\documentclass[prl,a4paper,twocolumn,10pt]{revtex4}
\usepackage{graphicx}
\usepackage{verbatim}
\usepackage{bm}

\begin{document}

\title{Theory of strong-field ionization of aligned CO$_2$}

\author{M. Abu-samha and L.~B. Madsen}
\affiliation{Lundbeck Foundation Theoretical Center for Quantum System Research, Department of Physics and Astronomy, Aarhus University, 8000 Aarhus C, Denmark.}

\begin{abstract}
A theoretical framework for studying strong-field ionization of aligned molecules is presented, and alignment-dependent ionization yields are computed for CO$_2$. Our calculations are in unprecedented agreement with recent experiments. We find that the ionization process is affected by intermediate resonance states, and the alignment-dependent ionization yields do not follow the electron density of the initial states. The theory explains the breakdown of semi-analytical theories, like the molecular tunneling theory and strong-field approximation, were excited electronic structure is neglected.
\end{abstract}

\maketitle

\begin{figure}
{\includegraphics[width=0.45\textwidth]{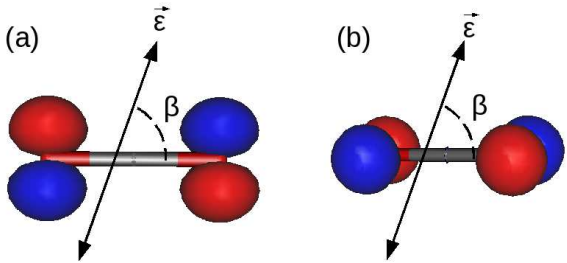}}
\caption{(Color online) Illustration of the orientation ($\beta$) of the two degenerate HOMO orbitals of CO$_2$ with respect to the linear polarization axis ($\vec{\varepsilon}$) of the laser field. The contribution to the ionization yield from (b) is very small in comparison with that from (a), and is not considered in the present study.}
\label{fig1}
\end{figure}

Strong-field physics emerges as a very promising field for studying the structure and dynamics of molecular systems. For example, recent developments in this field led to experiments on manipulating chemical reactions~\cite{Kling04142006}, tomographic imaging of molecular orbitals~\cite{nature04}, probing the nuclear dynamics on the attosecond~\cite{S.Baker04212006} and femtosecond~\cite{Science09} time scales, molecular alignment~\cite{PhysRevLett.85.2470} and torsional control~\cite{CBM09}.


Molecules can be aligned relative to the laser field, and in order to fully exploit this effect, it is necessary to understand the dependence of the initial ionization step on the molecular orientation. In~\cite{PhysRevLett.90.233003,PhysRevLett.93.113003,pavicic:243001,kumarappan093006}, alignment-dependent ionization yields were measured for ionization from the highest occupied molecular orbital (HOMO) of N$_2$ ($\sigma_g$(2p)), O$_2$ ($\pi_g$(2p)), CO$_2$ ($\pi_g$(2p)) and CS$_2$ ($\pi_g$(3p)) in the transition regime between tunneling and multiphoton ionization, and a strong dependence on the alignment angle ($\beta$ in Fig.~\ref{fig1}) was found. It was suggested that the angular dependence of the ionization directly maps to the orbital symmetry. The theoretical advance in the molecular case is impeded by the complexity of the molecular structure, which arises from the multicentre character of the molecular potential and the additional rovibrational degrees of freedom. Full \textit{ab initio} calculations of the alignment-dependent ionization are available only for H$_2^+$~\cite{0953-4075-42-2-021001,kamta,telnov:043412,kjeldsen:035402} and H$_2$~\cite{nikolopoulos:033402,awasthi:063403}. For larger molecules, despite a tremendous amount of experiments, no \textit{ab initio} calculations are available, and the most widely used approaches to explain strong-field processes are the molecular tunneling theory~\cite{PhysRevA.66.033402} and strong-field approximation~\cite{0953-4075-37-10-003,PhysRevLett.85.2280}. Calculations of alignment-dependent ionization yields based on these theories fail to explain recent experiments~\cite{pavicic:243001}: Tunneling theory and strong-field approximation predict the ionization yield to follow the electron density of the initial electronic state, in contrast with observations for the CO$_2$ molecule~\cite{pavicic:243001}.

\begin{figure}
{\includegraphics[width=0.35\textwidth]{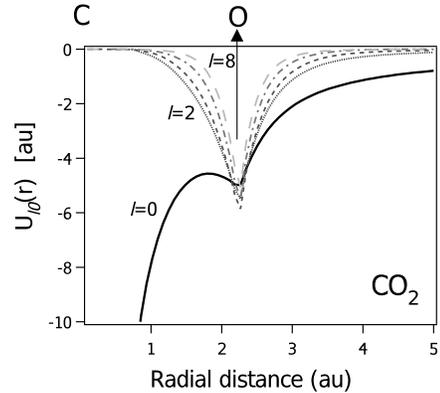}}
\caption{Radial molecular potentials (\ref{Eq2}) of CO$_2$ up to $l$=8.}
\label{fig2}
\end{figure}

In this Letter, we use \textit{ab initio} theory (within the single-active electron approximation) to investigate the response of linear, polyatomic molecules to intense femtosecond laser pulses, and in view of recent experimental results~\cite{pavicic:243001} we focus on the CO$_2$ molecule. The time-dependent Schr\"{o}dinger equation (TDSE) describing the active electron in the combined field of the frozen core and the laser pulse is solved numerically using grid methods~\cite{kjeldsenPRA07}, and the time-dependent wavefunction is analyzed using grid-based spectral methods~\cite{nikolopoulos:063426,PhysRevA.38.6000}. With this theory, we are able to explain many features in the experiments~\cite{pavicic:243001}, and clearly identify the short-comings of the tunneling theory~\cite{PhysRevA.66.033402} and the strong-field approximation~\cite{0953-4075-37-10-003,PhysRevLett.85.2280}. In short, the present work underlies the importance of dynamics in the excited state manifold, and extends studies within the single-active electron model to systems beyond H$_2^+$, H$_2$ and their isotopes.

The time-dependent wavefunction is solved in a partial wave expansion. Thus, it will be convenient if the potential describing the active electron ($V(\vec{r})$) is also expanded in partial waves. $V(\vec{r})$ is obtained from quantum mechanical calculations~\cite{GAMESS}, and electron exchange is treated within the local-density approximation~\cite{DFT-LDA}. The radial components of the potential, $U_{lm}(r)$, are obtained by integrating over the angular variables, i.e.
\begin{equation}
\label{Eq2}
U_{lm}(r)=\int d\Omega Y_{lm}^{\ast}(\Omega)V(\vec{r}).
\end{equation}
The integral is evaluated numerically using cubature on the sphere~\cite{PhysRevA.68.031401}. The radial potentials are shown for CO$_2$ in Fig.~\ref{fig2}. The potential provides accurate description of the HOMO orbital (both orbital energy and angular decomposition) and the excited states of the molecule; the energies of the first and second excited states with $\pi_u$ symmetry (5.85~eV and 2.72~eV) are in agreement with the values (4.75~eV and 2.48~eV) reported in x-ray absorption spectroscopy~\cite{GUNNELIN98} (see also Fig.~\ref{fig5}).

\begin{figure}
{\includegraphics[width=0.35\textwidth]{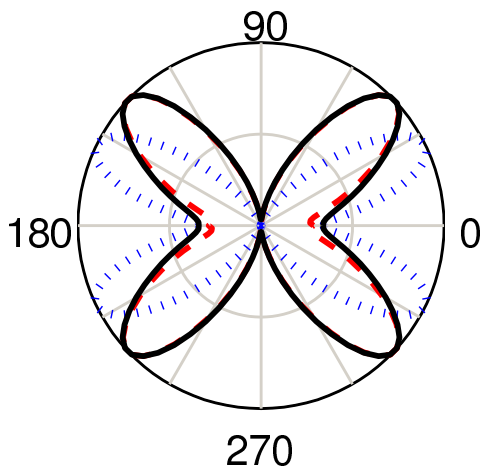}}
\caption{Ionization yields as a function of the angle $\beta$ (see Fig.~\ref{fig1}) for CO$_2$. The dashed (solid) line denotes calculations based on the present approach for a 10-cycle laser pulse at 800~nm and laser peak intensities 5.6$\times$10$^{13}$W/cm$^{2}$ (1.1$\times$10$^{14}$W/cm$^{2}$). The dotted line denotes calculations based on the molecular tunneling theory~\cite{pavicic:243001}. The computed yields are given relative to the orientation that gives the maximum yield.}
\label{fig3}
\end{figure}

\begin{figure}
{\includegraphics[width=0.5\textwidth]{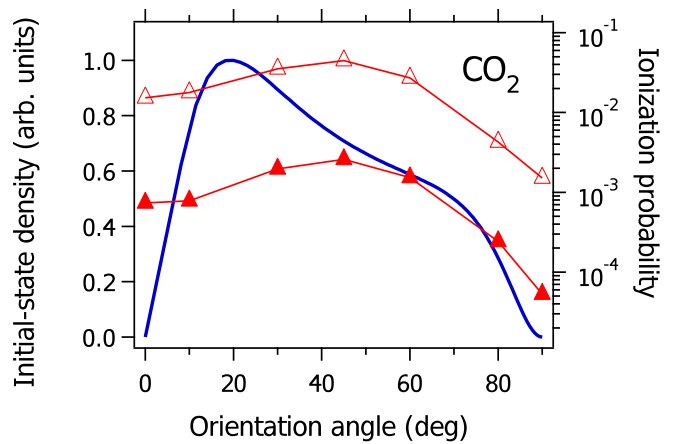}}
\caption{Initial-state electron density (thick, solid line; axis to the left) and ionization yields (triangles; axis to the right) in strong-field ionization of CO$_2$ at laser intensities 1.1$\times$10$^{14}$W/cm$^{2}$ (open triangles) and at 5.6$\times$10$^{13}$W/cm$^{2}$ (filled triangles). }
\label{fig4}
\end{figure}

\begin{figure}
{\includegraphics[width=0.5\textwidth]{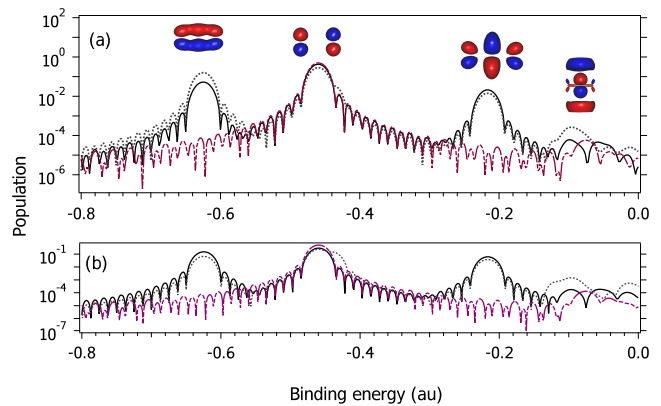}}
\caption{Bound-state energy spectrum of CO$_2$ as computed at the peak intensity of a 10-cycle, 800-nm pulse ($\vec{A}$(t)=0) with intensity (a) 5.6$\times$10$^{13}$W/cm$^{2}$ and (b) 1.1$\times$10$^{14}$W/cm$^{2}$. The solid, dotted and dashed lines, respectively, denotes $\beta$ values 0$^\circ$, 45$^\circ$ and 90$^\circ$. In (a), the spectral structure from left to right is assigned to the following electronic states: $\pi_u$ (HOMO-1),  $\pi_g$ (HOMO), $\pi_u$ (LUMO) and $\pi_u$ (C 3p). }
\label{fig5}
\end{figure}
The TDSE is solved on a grid within the velocity gauge~\cite{kjeldsenPRA07}. The external field, linearly polarized along $\vec{\varepsilon}$, is characterised by the vector potential $\vec{A}(t)=A_0\sin^2(\pi t/T)\cos(\omega t)\vec{\varepsilon}$ with $T$ the pulse duration, $\omega$ the frequency and $A_0=E_0/\omega$ with $E_0$ the field strength. We use an equidistant grid with 2048 points that extends up to 160~au. The laser pulses contain 10 cycles, and the calculations were performed at 800~nm and peak intensity 5.6$\times$10$^{13}$W/cm$^{2}$ and 1.1$\times$10$^{14}$W/cm$^{2}$. The angular basis set contains 21 spherical harmonics. The ionization yields are calculated from the numerical grid calculations by applying an absorbing boundary~\cite{nikolopoulos:063426}. The calculations were repeated in a larger box (320~au with 4096 grid points) with a larger angular basis (31 spherical harmonics) and the results are converged.

The ionization yields are shown in Fig.~\ref{fig3} as obtained from our TDSE calculations at 5.6$\times$10$^{13}$W/cm$^{2}$ and 1.1$\times$10$^{14}$W/cm$^{2}$, and from molecular tunneling theory~\cite{pavicic:243001}. Starting with the TDSE results, the yields are very sharp and peak at 45$\pm3^\circ$. The effect of laser intensity is rather negligible; it only enhances the ionization yield somewhat when the molecule is aligned parallel to the laser field. The molecular tunneling theory predicts the ionization yield to peak at $\beta$=25$^\circ$.

In Fig.~\ref{fig4}, we compare the ionization yields at orientation $\beta$, to the electron density of the HOMO orbital at that orientation. One clearly sees that the ionization yields do not follow the electron density of the initial state, e.g., the HOMO orbital density for CO$_2$ peaks at $\beta$=20$^\circ$ whereas the ionization yield peaks at 45$\pm 3^\circ$. This indicates that excited electronic states, accounted for in the present approach, contribute to the ionization process. To investigate this point further, we computed the bound states energy spectrum~\cite{PhysRevA.38.6000,nikolopoulos:063426}
\begin{equation}
\label{Eq3}
\frac{dP(E,\beta)}{dE}=\frac{1}{\tau} \left\vert \int_T^{T+\tau}dtC(t)e^{iEt}\mathbf{w}(t) \right\vert,
\end{equation}
where $\mathbf{w}(t)$ is a hanning function and $C(t)$ the autocorrelation function of the time-dependent wavefunction $\psi(t)$. The bound-state energy spectra are shown for CO$_2$ in Fig.~\ref{fig5}, as obtained from the calculations at 5.6$\times$10$^{13}$W/cm$^{2}$ and 1.1$\times$10$^{14}$W/cm$^{2}$. The energy spectra are similar for the two laser intensities; the spectra show that ionization occurs via several excited states, namely the lowest unoccupied molecular orbital (LUMO) at energy 0.21~au below the threshold, and the carbon $\pi_u$(3p) excited state at energy 0.10~au below the threshold. The excited states were identified by comparison to theoretical oxygen K-edge x-ray absorption spectra of CO$_2$~\cite{GUNNELIN98}. The population in the LUMO state depends strongly on $\beta$, and for $\beta$ values up to 80$^\circ$, it is significantly larger than the ionization yields. At 5.6$\times$10$^{13}$W/cm$^{2}$, the total ionization yields (LUMO population at the peak of the pulse; $\vec{A}$(t)=0) at $\beta$ values of 0$^\circ$, 45$^\circ$, 80$^\circ$ and 90$^\circ$, respectively, are 7.35$\times$10$^{-4}$~(2.01$\times$10$^{-2}$), 2.58$\times$10$^{-3}$~(1.34$\times$10$^{-2}$), 2.43$\times$10$^{-4}$~(1.33$\times$10$^{-3}$) and 5.31$\times$10$^{-5}$~(9.78$\times$10$^{-7}$). The energy spectrum shows that some of the HOMO population gets transferred into the HOMO-1, $\pi_u$(2p) orbital. A crude estimate of the relative contribution from this state to the total ionization yield is about 10\%, based on a simple model that takes into account the relative population and the difference between the ionization potentials of the HOMO and HOMO-1 orbitals. 

\begin{figure}
{\includegraphics[width=0.35\textwidth]{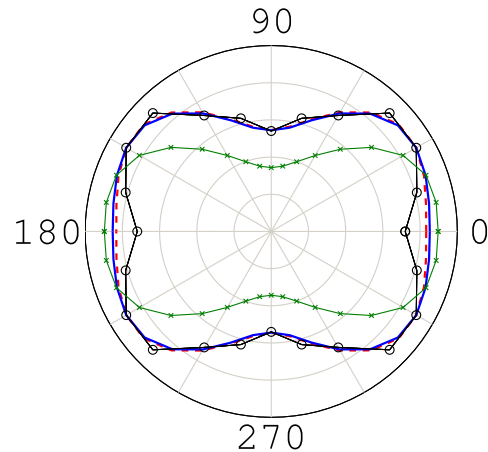}}
\caption{(Color online) Experimental (circles) and theoretical (dashed line: 5.6$\times$10$^{13}$W/cm$^{2}$; solid line: 1.1$\times$10$^{14}$W/cm$^{2}$; crosses: MO-ADK theory)~\cite{pavicic:243001} ionization yields for CO$_2$.}
\label{fig6}
\end{figure}

Now we compare our calculations with the experimental measurements for CO$_2$ at 1.1$\times$10$^{14}$W/cm$^{2}$~\cite{pavicic:243001}. First, it should be emphasized that while the present calculations are carried out at a fixed laser intensity, the measurements take into account focal volume effects, fluctuations of laser intensity and alignment distribution. Including focal volume effects and fluctuation of laser intensity places heavy demands on the computational resources, and since the ionization yields show little dependence on laser intensity (cf. Fig.~\ref{fig3}), these contributions will be left out. However, we do convolute our results with the alignment distribution function devised in~\cite{pavicic:243001}. The final results are shown in Fig.~\ref{fig6}. We also apply the same convolution procedure to the ionization yields obtained from the tunneling theory~\cite{pavicic:243001}. One can clearly see that the tunneling theory is no where near the measurements. By contrast, our TDSE calculations show excellent agreement with the measurements. This demonstrates the important role of excited states in strong-field ionization of CO$_2$.

In the preceding paragraphs, we demonstrated superiority of the proposed approach over the molecular tunneling theory and the strong-field approximation. This is because the present approach takes into account the molecular electronic potential, and hence is capable of describing the ionization process in different ionization regimes. The tunneling theory~\cite{PhysRevA.66.033402} and strong-field approximation~\cite{0953-4075-37-10-003,PhysRevLett.85.2280}, on the other hand, assume a pure ionization mechanism (pure tunneling or direct multiphoton ionization), without accounting for excited electronic structure. Therefore, they fail to explain the experimental measurements.

While the discussion above focused on the alignment-dependent ionization only, the present methodology can be extended to extract other observables of interest. For example, the above-threshold-ionization spectrum can be determined by extending the analysis based on the autocorrelation function to the continuum~\cite{nikolopoulos:063426}, the photoelectron distribution can be determined by flux analysis~\cite{KjeldsenPRA74} and the full momentum distribution can be determined by projection on exact scattering states~\cite{0953-4075-42-2-021001}, or by analysis based on asymptotic projection operators~\cite{MadsenPRA76}. Also, since the expectation of the dipole acceleration is readily evaluated from the time-dependent wave packet, the method can be extended to consider the process of high-harmonic generation in molecules. 

In conclusion, our theoretical calculations are in unprecedented agreement with recent experiments, and explain the breakdown of the tunneling theory and strong-field approximation. The main finding is that the excited electronic structure is vital in strong-field ionization of molecules, and the measured ionization yields do not follow the electron density of the initial state. The dynamics in the excited state manifold will affect not only the ionization dynamics but surely also the process of high-harmonic generation and in particular the mapping to the electronic continuum of the excited state symmetry may impede the extension of tomography techniques to more general classes of molecules. The proposed framework for treating polyatomic molecules is grid based, which is the most widely used approach in strong-field physics, and takes input potentials from standard quantum chemistry codes~\cite{GAMESS}. It could therefore be easily implemented by many research groups, and will facilitate studies of systems more complex than H$_2$ and H$_2^+$ and their isotopes. 

We thank H. Stapelfeldt for useful comments on the manuscript. This work was supported by the Danish Research Agency (Grant No. 2117-05-0081).


\end{document}